\documentclass{aa}  

\usepackage{hyperref}
\usepackage{graphicx}
\usepackage{soul}
\usepackage{txfonts}
\usepackage{amsmath}
\usepackage{gensymb}
\usepackage{xcolor}
\usepackage{subfig}
\usepackage{caption}
\usepackage{array}
\usepackage{siunitx}

\newcommand{\pkg}[1]{\texttt{#1}}

\makeatletter
\renewcommand{\fnum@table}{\textbf{Table~\thetable}}
\makeatother

\makeatletter
\renewcommand{\fnum@figure}{\textbf{Fig.~\thefigure}}
\makeatother
\usepackage{xcolor}

\usepackage{textcomp}

\begin{document}

   \title{Assembly bias and the redshift evolution of intrinsic alignments for LRGs}

  \author{A. Herle \inst{1}
  \and N. E. Chisari \inst{2, 1}
  \and H. Hoekstra \inst{1}
  \and D. Neumann \inst{1}
  }

   \institute{Leiden Observatory, Leiden University, Niels Bohrweg 2, 2333 CA, Leiden, the Netherlands\\
              \email{herle@strw.leidenuniv.nl}
         \and
              Institute for Theoretical Physics, Utrecht University, Princetonplein 5, 3584 CC, Utrecht, the Netherlands\\
             }

  \abstract 
   {The intrinsic alignment (IA) of galaxies is one of the main contaminants to the weak lensing shear signal. Efforts to model it often assume that the alignment strength depends only on halo mass. In this work, we use the 2.8 Gpc box-size run of the \pkg{FLAMINGO} suite to show that alignment amplitude, in addition to halo mass, depends on the formation redshift of the host haloes for an LRG-like sample. We show that the assembly histories of galaxies and haloes influence the alignment signal. After correcting for their mass evolution, we find that haloes that formed earlier have a higher alignment amplitude, as do their central galaxies. We also explore the redshift evolution of the alignment signal by fitting the amplitude with a power-law in mass at different snapshots of the simulation. We find that the amplitude of this power-law increases steadily with redshift, while the slope decreases with redshift until $z\sim1$ and then flattens. We provide an empirical mass-redshift intrinsic alignment model fit on the \pkg{FLAMINGO} simulation. Furthermore, by tracking central galaxies across snapshots, we show that the alignment signal changes with redshift beyond that associated with the change in mass, and that galaxies tracked from higher redshifts have a larger amplitude. Our results indicate that IA modeling in weak lensing surveys cannot have arbitrarily small prior ranges, and complicate the implementation of HOD-based alignment models for gravity-only simulations. They also provide simulation-based guidelines for a redshift evolution model of IA for use in observational studies.}


   \maketitle

\section{Introduction}

The deflection of light from distant galaxies due to the mass of intervening structures is one of the most sensitive cosmological probes in modern cosmology. Surveys like \textit{Euclid} \citep{Laureijs2011, Mellier2025} and the Legacy Survey of Space and Time \citep[LSST, ][]{LSST2009, Ivezic2019} will have unparalleled depth and area, and are poised to answer long-standing questions about the nature of dark matter and dark energy using weak gravitational lensing which imprints correlations onto the shapes of galaxies. These can be used to infer cosmological parameters, but the interaction of galaxies with the tidal field of the matter distribution can introduce additional correlations, called intrinsic alignments (IA). IA has been shown to contaminate weak lensing analyses \citep{Heavens2000, Catelan2001, Hirata2004} and bias cosmological parameters if modelled incorrectly \citep[e.g.][]{Hirata2007, Joachimi2011, Samuroff2024}.

Thus, it is vital to properly model the IA of galaxies in a weak lensing survey to prevent biasing cosmological results. Numerous works have studied the IA of galaxies, both in observations and simulations. Halo mass has been identified as the main driver of the strength of alignments. Studies of simulations, such as the Millenium and Millenium-2 simulations \citep{Schneider2012} and the Millenium XXL simulation \citep{Piras2018} showed a significant dependence of the alignment amplitude with halo mass. Observational studies using SDSS \citep{Joachimi2011, Singh2015}, KiDS \citep{Johnston2019, Fortuna2021b, Georgiou2025}, DES \citep{Samuroff2023} and PAUS \citep{Navarro-Girones2026} also found an increase in the alignment amplitude for more luminous galaxies, and hence more massive ones. The dependence of the alignment amplitude on halo mass has been used successfully to study IA \citep{Fortuna2021a} and also led to the use of the mass-dependent Non Linear Alignment (NLA) model, NLA-M, as the fiducial alignment model in the KiDS-Legacy analysis \citep{Wright2025}. Recently, \citet{Herle2026} introduced the TATT-M model where the halo mass-dependence of the higher-order Tidal Alignment Tidal Torquing (TATT) terms were used to reduce the parameter space of TATT.

Many modeling efforts to date effectively assume that the alignment signal depends only on the mass of the halo. Given the observational uncertainties of Stage-III surveys like KiDS, any other dependencies beyond the halo mass are not detectable. Given the large statistical power of surveys like \textit{Euclid} and LSST, dependencies that were hidden in the error bars in previous surveys will become more relevant. This necessitates the use of ever-larger simulations to test these fundamental assumptions of the IA models used in cosmic shear surveys. One such dependence that is not accounted for in IA models in the literature, is the effect of baryonic physics. \citet{Herle2026} showed that once the effect of feedback on the halo mass is accounted for, feedback itself does not change the alignment amplitude within the error bars of \pkg{FLAMINGO} \citep{Schaye2023}. However, there is still another effect that may influence the alignment signal, namely, the formation redshift.

According to the linear alignment model \cite{Hirata2004}, intrinsic alignments are imprinted at the time of halo (or galaxy) formation. Works that have studied the alignment bias parameters in $N$-body simulations have indeed confirmed they satisfy co-evolution relations consistent with this assumption, also termed the ``Lagrangian prior'' \citep{Bakx2023,Vedder2026}. \citet{Camelio2015} and more recently \citet{Maion2025}, argued that the mass-dependence of linear intrinsic alignment bias is too strong to be the result of post-formation mechanisms. Therefore, it is clear that alignments must be imprinted at some formation redshift \citep[see also][]{Schmitz2018}.

This dependence of the alignment amplitude on formation redshift does not preclude a dependence on mass. In fact, \citet{Piras2018} and \citet{Maion2025} argue in favour of such a dependence, inherited from the response of the quadrupolar shape of the object on the variance of the local tidal field. This mass-dependence can also be expressed as a dependence on matter density field peak height: the stronger the clustering in the region where the object forms, the stronger the variance of the tidal field and the corresponding alignment. Still, it remains unclear whether the intrinsic alignment amplitude depends on this quantity only, or whether secondary dependencies are also present. In the clustering literature, a similar discussion on the dependence of halo (or galaxy) properties on other quantities beyond mass goes under the name of ``assembly bias'' \citep{Sheth2004}. Investigations of this phenomenon in the context of galaxy clustering can be found \citep[for example][]{Zhu2006, Croton2007} and they reveal that objects that formed earlier systematically lie in more overdense regions than those that formed later.

\citet{Sheth2004} used early gravity-only simulations to show that haloes that formed at higher redshift are clustered more strongly than later forming haloes. This was further corroborated by later works using the Millenium simulations, which showed that clustering depends strongly on the halo formation time, and that halo mass was not the sole driver of clustering  \citep{Gao2005, Croton2007, Jing2007}.  \citet{Wang2013} provided the first observational detection of assembly bias, as they showed that SDSS central galaxies with lower specific star formation rate (sSFR; i.e older) cluster more strongly than those with higher sSFR. \citet{Lacerna2014} applied their halo mass-independent definition of age \citep{Lacerna2011} to divide the SDSS sample into older and younger galaxies, and found a significant dependence of the clustering amplitude with age.

Tentative evidence for the impact of assembly bias on intrinsic alignments was found in \citet{Maion2025}, via a secondary dependence of the linear alignment parameter on halo spin. In addition, halo shapes are known to be subject to assembly bias, with for example \citet{Maccio2007} and \citet{Schneider2012} finding that haloes that are more spherical formed earlier. In this work, we explicitly look for secondary dependencies of the alignment amplitude on the formation redshift beyond halo mass.

Understanding the impact of assembly history on the IA signal also requires an understanding of how the alignment of haloes and galaxies changes over cosmic time. The redshift evolution of the alignment signal has also garnered a lot of interest in the IA literature over the years \citep[see][for a review of this topic]{Chisari2025}. Furthermore, an IA model that accounts for the evolution of the alignment amplitude with both mass and redshift is very useful in the context of weak lensing analyses. A redshift-dependent NLA model (NLA-$z$) was tested in the KiDS Legacy analysis \citep{Wright2025}, where the alignment amplitude was allowed to increase with redshift. However, a redshift scaling has yet to be detected in observations \citep[see for example][and references therein]{Navarro-Girones2026}, but this is possibly a consequence of the large uncertainties in cosmic shear samples \citep[which was also noted by][in their figure 10]{Abbott2026}. This motivates the use of large hydro-dynamic simulations to understand the redshift evolution, if any, of the alignment signal.

In this work, we use the \pkg{FLAMINGO} simulations to study the impact of assembly history on IA and introduce an empirically derived mass-redshift IA model. In Section \ref{sect:formalism} we introduce the theoretical background for this work, and in Section \ref{sect:flamingo} we describe the \pkg{FLAMINGO} simulation run we used. We introduce our definition of formation redshift and explore the dependence of the alignment signal on it in Section \ref{sect:zf_dependence}. We present a mass-redshift model for the intrinsic alignment amplitude and apply it to a consistent sample of galaxies tracked across snapshots in Section \ref{sect:zevo}.

\section{Formalism}
\label{sect:formalism}

The formalism used in this work is identical to that in \citet{Herle2026} and we refer the reader to their section 2. In this section, we introduce some of the main aspects of the formalism. 

\subsection{Estimators}
In order to calculate the position-position (`gg') and position-shape (`g+') correlation functions that we will use in this work, we employ the generalised Landy-Szalay correlation function estimators \citep{Landy1993}, which are defined as:

\begin{equation}
    \xi_{\mathrm{gg}} = \frac{(S - R_S)(D - R_D)}{R_S R_D} = \frac{SD - R_SD - SR_D + R_S R_D}{R_S R_D},
\end{equation}

\noindent and 

\begin{equation}
    \xi_{\mathrm{g}+} = \frac{S_+D - S_+R_D}{R_S R_D},
\end{equation}

\noindent where $S$ is the shape sample and $D$ is the position sample. $R_S$ and $R_D$ are the shape and position random samples respectively. For all measurements in this work, we employ a random catalogue that is ten times larger than the dataset. We verified convergence by comparing with a catalogue twenty times larger and found that the increase does not change our results. The terms containing shears are calculated using

\begin{equation}
    S_+X = \sum_{i\in S, j\in X} \gamma_+^{(i)}\langle j | i \rangle.
\end{equation}
\noindent where $X$ is either $D$ or $R_D$, $\gamma_+$ measures the components of the shear along the line joining the pair of galaxies, and  $\langle j | i \rangle$ indicates galaxy shape $i$ with respect to separation vector towards galaxy $j$.

All correlation functions presented in this work were computed using \pkg{IACorr},\footnote{https://github.com/elizabethjg/IACorr} which is built upon \pkg{TreeCorr} \citep{Jarvis2004}. We used the delete one jack-knife (JK) method to estimate covariance matrices \citep[for more details, see][]{Norberg2009}. The covariance matrix is given by

\begin{equation}
\mathrm{Cov}_{\mathrm{JK}} =
\frac{N_{\mathrm{JK}}-1}{N_{\mathrm{JK}}}
\sum_{i=1}^{N_{\mathrm{JK}}}
\bigl(
\mathbf{w}_{\mathrm{ab},i}-\bar{\mathbf{w}}_{\mathrm{ab}}
\bigr)
\bigl(
\mathbf{w}_{\mathrm{ab},i}-\bar{\mathbf{w}}_{\mathrm{ab}}
\bigr)^{\mathrm{T}},
\end{equation}

\noindent where $N_{\mathrm{JK}}$ is the number of jack-knife patches, $\mathbf{w}_{\mathrm{ab},i}$ is the correlation function after removing the signal of the $i$th JK region and $\bar{\mathbf{w}}_{\mathrm{ab}}$ is the mean of all the JK regions. We use $N_{\mathrm{JK}} = 64$ for all the error bars in this paper. Typically, a higher number of patches is used, but given that we have smaller samples after binning in mass, we compromise and use 64 patches per box.

We jointly modelled $w_{\mathrm{gg}}$ and $w_{\mathrm{g+}}$, as described in the next section. To model our data vectors, we used \pkg{IATheory} \footnote{https://github.com/DavidNavarroG/IATheory} \citep{Navarro-Girones2026, Herle2026}. We use this code to model the data vectors using the {\tt nautilus} sampler \citep{nautilus}.

\subsection{Theoretical background}

The projected position-position (`gg') and position-shape (`g+') correlation functions \citep{Blazek2011, Singh2016} can be expressed as

\begin{align}
\label{eqn:wgg}
w_{\mathrm{gg}}(r_{\mathrm{p}}) &= \frac{1}{\pi^2} 
\int_0^{\infty} \mathrm{d} k_z 
\int_0^{\infty} \mathrm{d} k_{\perp} \, 
\frac{k_{\perp}}{k_z} \,\nonumber\\
&\quad \times P_{\mathrm{gg}}(\vec{k}, z_{\mathrm{s}})  \sin(k_z \Pi_{\mathrm{max}}) \, J_0(k_{\perp} r_{\mathrm{p}}),
\end{align}

\noindent and 

\begin{align}
\label{eqn:wgp}
w_{\mathrm{g+}}(r_{\mathrm{p}}) &= \frac{1}{\pi^2}
\int_0^{\infty} \mathrm{d} k_z 
\int_0^{\infty} \mathrm{d} k_{\perp} \,
\frac{k_{\perp}^3}{k_{\perp}^2 + k_z^2} \, \nonumber\\
&\quad \times P_{\mathrm{gI}}(\vec{k}, z_{\mathrm{s}}) \sin(k_z \Pi_{\mathrm{max}}) \, J_2(k_{\perp} r_{\mathrm{p}}),
\end{align}

\noindent where $r_{\mathrm{p}}$ is the 2D separation, $\Pi_{\mathrm{max}}$ is the maximum line-of-sight separation integrated over, $k$ is the wavenumber, $J_0$ and $J_2$ are cylindrical Bessel functions of the first kind, of order 0 and 2 respectively, $k_{\perp}$ is the perpendicular wavenumber, $k_z$ is the wavenumber in the $z$ direction, and $z_{\mathrm{s}}$ is the redshift of the simulation snapshot. $P_{\mathrm{gg}}$ denotes the auto-correlation of galaxy positions, and $P_{\mathrm{gI}}$ quantifies the correlation between the galaxy positions and the intrinsic ellipticities.

At large scales, $P_{\mathrm{gg}}$ can be related to the density field with a linear-bias parameter \citep{Kaiser1984}, but this can be extended to smaller scales using the non-linear expansion 

\begin{align}
P_{\mathrm{gg}}(k) &= b_1^2P_{\delta \delta}(k) + b_1 b_2P_{b_1 b_2}(k) \nonumber\\
&\quad + b_1 b_{s^2}P_{b_1 s^2}(k) + b_1 b_{3\mathrm{nl}} P_{b_1 b_{3\mathrm{nl}}}(k) \nonumber\\
&\quad + \tfrac{1}{4}b_2^2P_{b_2 b_2}(k)
 + \tfrac{1}{2}b_2b_{s^2}P_{b_2 s^2}(k)
 + \tfrac{1}{4}b_s^2P_{s^2 s^2}(k),
\end{align}

\noindent where $s$ is the tidal field, $P_{\delta \delta}$ is the non-linear matter power spectrum and the power spectrum kernels $P_{b_1}$, $P_{b_2}$, $P_{b_1s^2}$ etc. are defined in \citet{Saito2014}. We also invoke the co-evolution relations $b_{s^2} = -\frac{4}{7} (b_1 -1) $ and $b_{3\mathrm{nl}} = b_1 -1$ to reduce our parameter space \citep[][see Section \ref{sect:zf_dependence} for a discussion on the impact of this assumption on our results]{Saito2014}. Finally, $P_{\mathrm{g+}}$ can be expressed as 

\begin{equation}
    P_{\mathrm{g+}}(k, z) = b_1 P_{\delta+}(k, z),
\end{equation}

\noindent where $P_{\delta+}$ is the matter-intrinsic power spectrum. In this paper, we use the Non-linear Alignment (NLA) model of IA, as motivated below.

The linear alignment model \citep{Catelan2001, Hirata2004} assumes that the alignments of galaxies are imprinted at the time of galaxy formation by the initial tidal field of the galaxy's environment. Since this model stems from linear theory alone, it performs well primarily on  large scales ($>$ 100 Mpc). This model originally used the linear matter power spectrum, but using the non-linear matter power spectrum instead, as proposed by \citet{Hirata2004} and implemented by \citet{Bridle2007}, this can be extended to quasi-linear scales. This is known as the NLA model. Following this model,

\begin{equation}
    P_{\delta +}(k, z) = C_1(z) P_{\delta \delta}(k, z),
\end{equation}

\noindent with 
\begin{equation}
\label{eqn:a1_norm}
    C_1(z) = - A_1 \bar{C_1} \rho_{\mathrm{crit}} \Omega_{\mathrm{m}}D(z)^{-1},
\end{equation}

\noindent where $A_1$ is the IA amplitude, $\rho_{\mathrm{crit}}$ is the critical density, $\Omega_{\mathrm{m}}$ is the matter density fraction and $D(z)$ is the linear growth factor normalised such that it is 1 at $z = 0$. 

The NLA model has been shown to perform well on intermediate scales in both observational and simulation-based studies \citep{Fortuna2025, Paviot2026}. It involves fitting two parameters: the galaxy bias $b_1$ and the intrinsic alignment amplitude $A_1$. The position-shape correlation function $w_{\mathrm{g+}}$ constrains the product $A_1b_1$, while joint modelling with the clustering correlation function $w_{\mathrm{gg}}$, which constrains $b_1$, breaks the degeneracy between these parameters.

More complex models that include higher order terms than we will consider here exist, for example the TATT model. NLA is typically the simplest model used in practice, and can be interpreted as a subset of the TATT model, with only $A_1$ allowed to be non-zero. TATT has been shown to represent the \pkg{FLAMINGO} data better than NLA due to the inclusion of higher-order terms \citep{Herle2026}, although the velocity-shear term has been shown to be very important as well \citep[see][]{Vedder2026}. Since we are interested in showing the impact of formation time on the alignment amplitude of galaxies, we restrict this work to the NLA model, and only use scales at which this model works well i.e 10 - 100 Mpc/$h$ \citep{Herle2026}. If assembly bias affects the alignment amplitude inferred with NLA, higher-order models should also be affected, and showing its impact within NLA for our conservative scale cuts would constitute a sufficient test.

\section{\pkg{FLAMINGO} simulations}
\label{sect:flamingo}

Throughout this paper, we used the output of \pkg{FLAMINGO}, which is a public Virgo consortium project presented in \citet{Schaye2023} and \citet{Kugel2023}, with the data release outlined in \citet{Helly2026}. \pkg{FLAMINGO} is a large suite of cosmological structure formation simulations that encompass variations in cosmology, baryonic feedback and numerical resolution. The simulations were performed with the SWIFT hydrodynamics code \citep{Schaller2024}. For details regarding the neutrino treatment and implementations of radiative processes, star formation, stellar mass loss and enrichment, stellar and AGN feedback and black-hole growth, we refer the reader to the overview paper \citep{Schaye2023}. 

In this paper, we make use of the 2.8 Gpc box-size run. The effect of resolution on halo shapes and orientations was explored by \citet{Herle2025}, and in this work we use only the runs with the fiducial resolution of $\mathrm{m}_{\mathrm{CDM}} = 5.65 \times 10^9 \ \mathrm{M}_{\odot}$. This is a $\Lambda$CDM run with mean CDM particle mass $\mathrm{m}_{\mathrm{CDM}} = 5.65 \times 10^9 \ \mathrm{M}_{\odot}$ and the number of baryonic particles $N_\mathrm{b}$ is 5040$^3$. The cosmology assumed is $h$ = 0.681, $\Omega_{\mathrm{m}} = 0.306$, $\Omega_{\Lambda} = 0.694$ , $\Omega_\mathrm{b} = 0.0486$,  $\Sigma \mathrm{m_\nu} \mathrm{c}^2 = 0.06$eV, $\mathrm{A}_\mathrm{s} = 2.099 \times 10^{-9}$, $\mathrm{n}_{\mathrm{s}} = 0.967$, $\sigma_8 = 0.807$, $\Omega_{\nu} = 1.39\times 10^{-3}$. This run uses the fiducial feedback variation of the \pkg{FLAMINGO} suite. The sub-grid prescriptions of stellar and AGN feedback were calibrated to the observed low redshift cluster gas fractions and galaxy stellar mass function, as described by \citet{Kugel2023}. 

Halo finding is a crucial step in analysing cosmological simulation data, and the choice of halo finder has been shown to influence results strongly \citep{Knebe2013a, Knebe2013b,  Moreno2025}. A 3D Friends-of-Friends (FoF) algorithm with linking length $l = 0.2$ times the mean CDM interparticle separation was first used to group particles using only the dark matter particles. Gas and stellar particles are then attached to the nearest CDM particle. \pkg{HBT-HERONS} \citep[Hierarchical Bound Tracing - Hydro-Enabled Retrieval of Objects in Numerical Simulations, ][]{Moreno2025}, which is a modified version of HBT+ \citep{Han2018}, then tracks haloes across timesteps to produce a subhalo catalogue. We used \pkg{HBT-HERONS} to produce subhalo catalogues, since it performs better than other subhalo finders \citep{Moreno2025} with better tracking of objects across snapshots, lower miscentering errors and better overall subhalo identification. We used catalogues of simulation quantities calculated by Spherical Overdensity and Aperture Processor (SOAP) \citep{McGibbon2025}.
\subsection{Inertia tensors and shapes}

\begin{figure}
    \centering
    \includegraphics[width=\columnwidth]{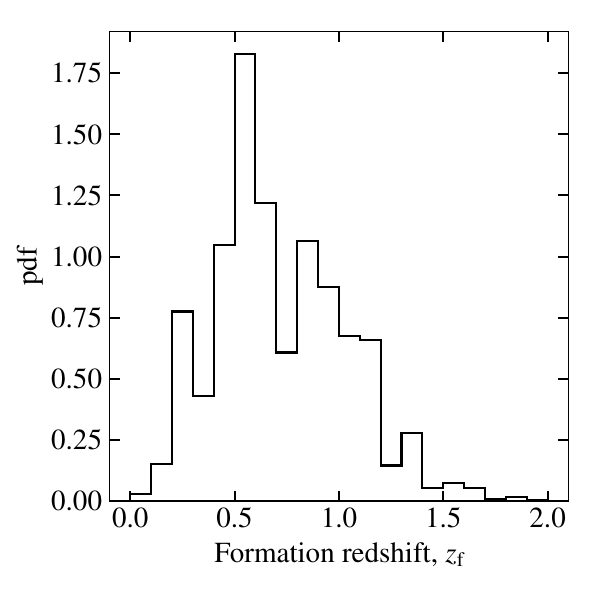}
    \caption{The distribution of formation redshifts, $z_{\mathrm{f}}$, for the sample of haloes used in this work. We defined $z_{\mathrm{f}}$ as the redshift corresponding to the snapshot at which the halo contained half of its total mass ($M_{\mathrm{200mean}}$) at present day.}
    \label{fig:zf_pdf}
\end{figure}

To estimate the shape of a given halo, we calculated its inertia tensor. Haloes were modelled as 3D ellipsoids whose axes point in the direction of the eigenvectors of the simple inertia tensor (SIT) defined as:

\begin{equation}
\label{eqn:sit}
    I_{ij} = \frac{1}{M}\sum_n m_{(n)}x_{i}^{(n)}x_{j}^{(n)},
\end{equation}

where $i, j = {1, 2, 3}$ correspond to the three axes of the simulation box, $m_n$ is the mass of the $n$th particle, $x_{i, j}^{(n)}$ are the positions of the $n$th particle in the $i$ or $j$ direction and $M$ is the total mass of the object. These were calculated for all bound particles within the half-mass radius of the objects using SOAP \citep{McGibbon2025}. We use the simple iterative tensor scheme, where there is no radial weighting of particles, and the shapes are calculated while iterating over the shape of the bounding ellipsoid. The choice of inertia tensor changes only the amplitude of the measured signal, and we refer the reader to appendix A of \citet{Herle2026} for a detailed comparison of how the choice of inertia tensor scheme affects results. Thus, our choice of the simple iterative tensor scheme over the reduced and non-iterative variations is motivated by the fact that it produces the strongest alignment amplitude and therefore has the highest signal-to-noise ratio for alignment studies. 

\begin{figure}
    \centering
    \includegraphics[width=\columnwidth]{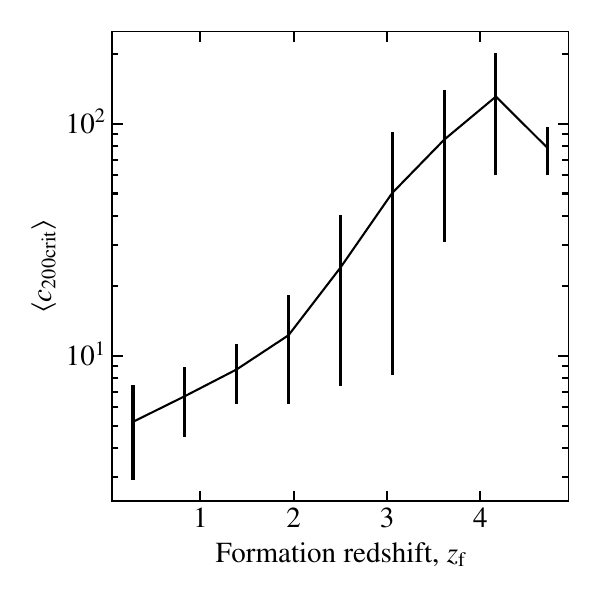}
    \caption{The mean concentration, $\langle c_{\mathrm{200crit}} \rangle$, for bins of formation redshift $z_{\mathrm{f}}$. The error bars on the y-axis correspond to the standard deviation of $c_{\mathrm{200crit}}$ in that $z_{\mathrm{f}}$ bin. We see that higher formation redshifts correspond to higher average concentrations, which motivates the use of this definition of $z_{\mathrm{f}}$.}
    \label{fig:c_vs_zform}
\end{figure}

The eigenvalues of the inertia tensor in Eqn. \ref{eqn:sit} are denoted $\lambda_i$, where $i = {1, 2, 3}$ correspond to the three axes of the ellipsoid. The lengths of the three axes of this ellipsoid, the major, intermediate and minor axes, are given by $a = \sqrt{\lambda_1}$, $b = \sqrt{\lambda_2}$ and $c = \sqrt{\lambda_3}$ respectively. Projected shapes are obtained by summing over only $i, j = 1,2$ in Eqn. \ref{eqn:sit}. We further define the axis ratio $q = b/a$. The eigenvectors of the inertia tensor, $\hat{e}_i$, encode the orientation of the object. 

\section{Dependence of $A_1$ on formation redshift}
\label{sect:zf_dependence}

In this section, we will explore how the alignment signal depends on halo formation time. We aim to compare both the halo and the galaxy alignment, so we need a consistent sample for which both the dark matter and stellar shapes are well resolved. Therefore, our halo sample consists of central haloes that have a bound dark matter mass more than $10^{12} \ \mathrm{M}_{\odot}$ and whose galaxies (which constitute our galaxy sample) have more than 300 stellar particles, corresponding to a stellar mass of $\sim 2.22 \times 10^{11} \ \mathrm{M}_{\odot}$. These cuts ensure that both the dark matter and stellar shapes we use in this work are unbiased; they have also been used extensively previously in the literature \citep{Chisari2015, Chisari2015b, Velliscig2015a, Herle2025, Herle2026}. 

Haloes form hierarchically, and their mass grows via accretion and mergers. In the literature, the formation redshift is often defined as the redshift at which a halo has acquired half of its final mass \citep{Sheth2004, Gao2005}. In Fig.~\ref{fig:zf_pdf}, we show the distribution of formation redshifts, $z_{\mathrm{f}}$, for our sample using this definition. Most haloes already possessed half their redshift 0 mass at $z \sim 1$, with some haloes having $z_{\mathrm{f}} \sim 2$. 

The concentration of an object when an NFW profile \citep{Navarro1997} is fit to it is often used as a proxy for formation time. Haloes that form earlier do so in a more dense universe than later forming haloes, making them more concentrated. This, coupled with the fact that they experience more fly-bys from satellites and companions, which preferentially strip the loosely bound outer material of the halo, results in a more concentrated object. This implies that any definition of $z_{\mathrm{f}}$ should correlate strongly with the concentration. In Fig.~\ref{fig:c_vs_zform}, we show the mean concentration, $c_{\mathrm{200crit}}$, in bins of $z_{\mathrm{f}}$. We see a clear correlation between $z_{\mathrm{f}}$ and concentration, which shows the validity of our definition. We tested another definition of formation redshift, where we defined it as the redshift corresponding to the snapshot at which the object first had more than 20 particles. This definition originates from simulation specific halo finding considerations, and we found that the formation redshift under this definition is not correlated with the concentration. In this work, we therefore also define the redshift corresponding to the snapshot at which the object possessed $0.5 M_{\mathrm{200mean}} (z=0)$ as the formation redshift.

\begin{figure}
    \centering
    \includegraphics[width=\columnwidth]{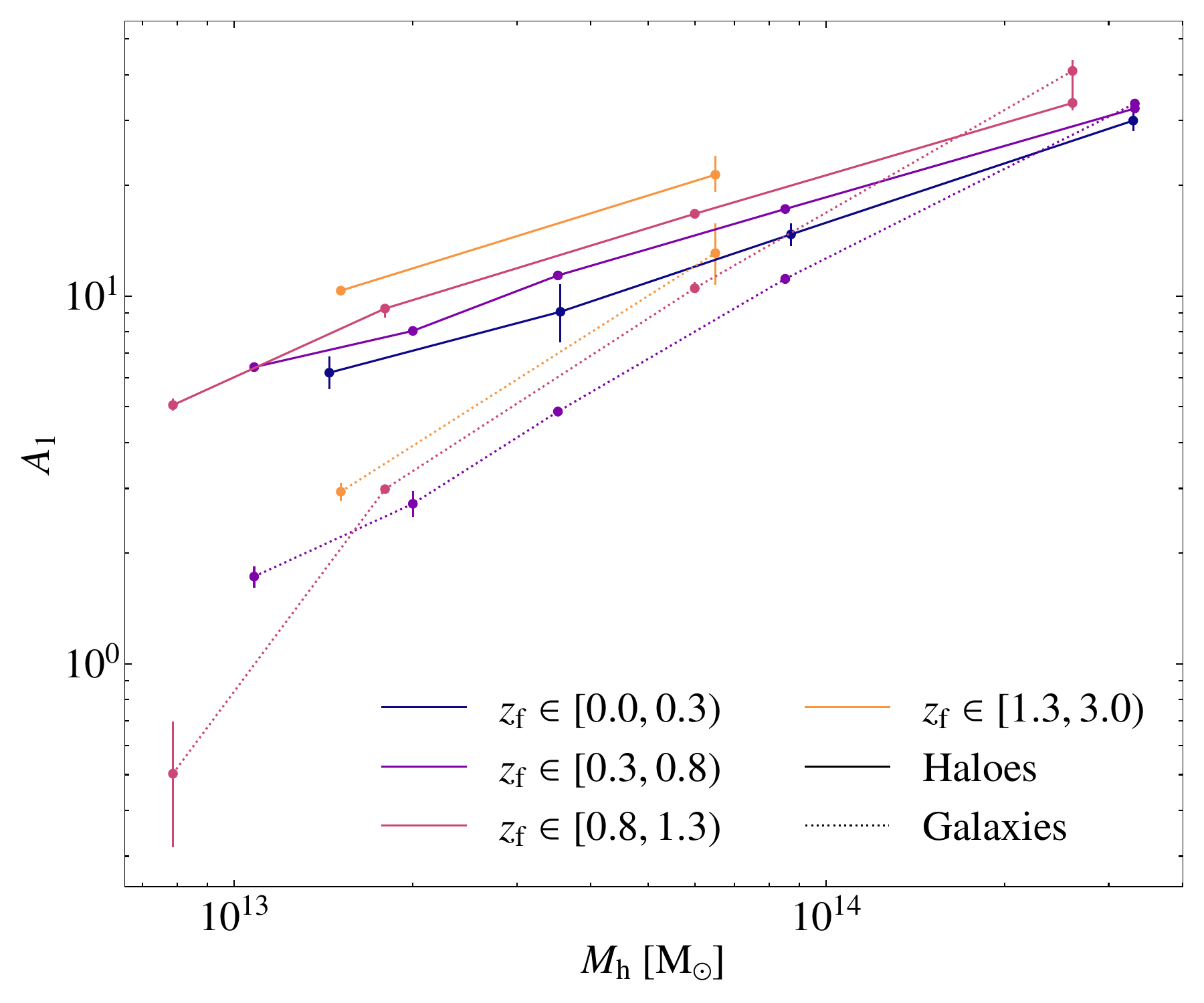}
    \caption{Variation of alignment amplitude $A_1$ with halo mass, $M_{\mathrm{h}}$. Here halo mass corresponds to the total $M_{\mathrm{200mean}}$ of the haloes of each object. We selected galaxies from the redshift 0 snapshot that had more than 300 stellar particles, and whose dark matter haloes have a bound dark matter mass $> 10^{12} \ \mathrm{M}_{\odot}$. The dotted lines indicate the galaxy alignment amplitudes, and the bold lines correspond to the halo alignments for the same sample. There is a clear increase of $A_1$ with halo mass. Moreover, there is a trend of increasing $A_1$ at the same halo mass for objects that formed earlier. }
    \label{fig:a1_vs_mh_zform}
\end{figure}

In the absence of assembly bias, one should be able to parametrise the alignment amplitude solely using mass. Before we test for secondary dependencies, we bin our sample of haloes at redshift 0 into bins of halo mass. For the halo mass, we used the total $M_{\mathrm{200mean}}$ of the haloes, which we refer to as $M_{\mathrm{h}}$. Within each mass bin, we further subdivided the sample according to $z_{\mathrm{f}}$. For every resulting ($z_{\mathrm{f}}$, $M_{\mathrm{h}}$) subsample, we measured the position-position and position-shape correlations and modelled them using the non-linear bias and NLA framework (described in Section \ref{sect:formalism}). 

The dependence of the alignment amplitude $A_1$ with halo mass is shown in Fig.~\ref{fig:a1_vs_mh_zform} for subsamples with different formation redshifts. The dotted lines correspond to the central galaxy alignment amplitudes, and the bold lines correspond to the $A_1$ of the haloes of those galaxies. Note that since the definition of formation redshift is based on the halo mass, for the lowest formation redshift bin the galaxy shapes are not yet converged and we exclude that bin for the galaxies. At a given halo mass, galaxies that formed earlier have a higher alignment amplitude. Looking at the haloes of these galaxies, we see that they too have a higher alignment amplitude for higher formation redshift objects. This similarity between the haloes and galaxies is expected given that galaxies tend to inherit their alignment from their host haloes. 

Moreover, later-forming objects have a higher mean halo mass than the full sample, whereas earlier-forming objects have a lower mean halo mass. If halo mass alone determined the alignment amplitude, these differences would simply move the subsamples along the $A_1$-$M_{\mathrm h}$ relation. Instead, Fig.~\ref{fig:a1_vs_mh_zform} shows that, at fixed halo mass, the relation is systematically offset with formation redshift, indicating that $z_{\mathrm f}$ introduces an additional dependence beyond halo mass.

The similarity in the slopes across bins of formation redshift is of note. Although the alignment amplitude is higher at each halo mass for objects that formed earlier, the slopes of the power-law remain unchanged. We find that the slopes of the power-law of haloes and galaxies are different, with galaxies having a steeper slope, with the most massive galaxies approaching the alignment amplitude of the most massive haloes.

To reduce the parameter space, we have invoked the co-evolution relations derived from the Lagrangian prior \citep{Saito2014}. There has been evidence in the literature, however, that assembly bias affects these co-evolution relations. \citet{Lazeyras2021} used N-body simulations to show that the $b_{s^2}(b_1)$ relation is significantly affected by assembly bias (their figure 5) which has also been confirmed in hydro-simulations, as \citet{Zennaro2022} showed that this relation has an increased scatter due to assembly bias. The concern then is that the breaking of these relations could bias the inferred alignment amplitude and lead to a spurious detection of the impact of assembly bias on IA. We confirm that this is not the case, by running our sampling procedure without these relations which does not change our inferred alignment amplitude and Fig.~\ref{fig:a1_vs_mh_zform} is unchanged. Without these relations, there are more bi-modalities in the posteriors of the clustering biases, but these have minimal impact on the inferred $A_1$. We also confirmed that $b_1$ is affected by assembly bias and the magnitude of the effect is consistent with the literature \citep{Gao2005}.

We also parametrised the alignment amplitude in terms of the peak height, following the approach of \citet{Maion2025}, to investigate the presence of assembly bias. We compared our results with theirs and found qualitative agreement. In particular, the shape-bias parameter $b_k$ exhibits a similar dependence on peak height $\nu$ when the sample is split into formation-redshift quartiles, analogous to the trend reported by \citet{Maion2025} for halo spin. A detailed presentation of these results is provided in appendix~\ref{appendix:peak_height}.

Fig.~\ref{fig:a1_vs_mh_zform} clearly demonstrates that the alignment strength depends not only on halo mass but also on formation redshift. The effect of formation redshift is primarily to shift the amplitude of the signal, while leaving its slope largely unchanged. This has important implications for IA modelling in cosmic shear analyses. Our results suggest that existing IA models do not require modification to account for assembly bias at the level of the functional form; however, assembly bias does impose a minimum uncertainty on the alignment amplitude, preventing arbitrarily narrow priors on this parameter.

These findings also have consequences for halo occupation distribution (HOD)-based approaches that assign galaxy alignments in gravity-only simulations \citep[for example][]{Castander2025}. Such methods will not reproduce the observed dependence on formation redshift unless this is explicitly incorporated into the modelling. This is particularly relevant for understanding the redshift evolution of the IA signal, which we investigate in the next section. Finally, we note that, given the mass resolution of the simulation, our conclusions apply only to the high-mass halo population associated with LRGs.

\section{Redshift evolution}
\label{sect:zevo}

\begin{figure}
    \centering
    \includegraphics[width=\columnwidth]{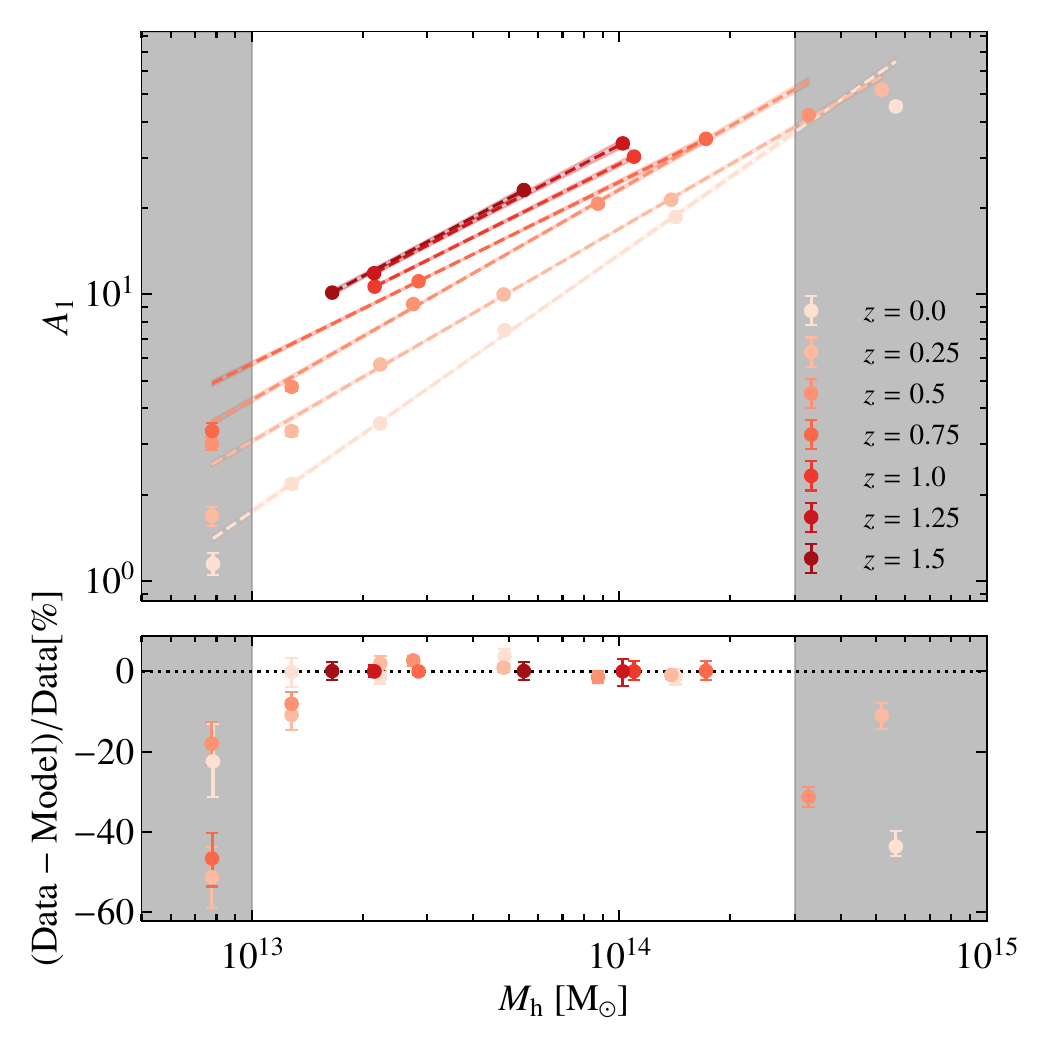}
    \caption{The variation of the alignment amplitude with halo mass at different redshifts. $A_1$ increases as a power-law of mass (Eqn. \ref{eqn:mass}). At higher redshifts, the overall amplitude at each mass is larger. In the bottom panel, we show the residuals with the best-fit power-law model. The grey regions indicate masses that are excluded from the fits.}
    \label{fig:a1_vs_mh_z}
\end{figure}

In the previous section, we showed that the alignment amplitude at a given halo mass depends on the formation redshift. In this section, we explore this effect using a different approach. In Fig.~\ref{fig:a1_vs_mh_z}, we show the variation of alignment amplitude with halo mass at different redshifts. This is different from Fig.~\ref{fig:a1_vs_mh_zform} where we show this variation at redshift 0, but in bins of $z_{\mathrm{f}}$. Instead here, at each redshift, we selected galaxies with more than 300 stellar particles, and whose dark matter haloes have a bound dark matter mass $> 10^{12} \ \mathrm{M}_{\odot}$. These objects are binned in halo mass, and an alignment amplitude is measured in each bin.

We see a clear increase in $A_1$ with halo mass, with galaxies at higher redshifts having a higher amplitude at a fixed mass. For each redshift, we fit a power-law between the amplitude and the mean halo mass \citep{Fortuna2025}:

\begin{equation}
\label{eqn:mass}
    A_1(M) = \alpha_{\mathrm{M}} \bigg( \frac{M}{M_0} \bigg)^{\beta_{\mathrm{M}}},
\end{equation}

\noindent where $\alpha_{\mathrm{M}}$ and $\beta_{\mathrm{M}}$ are the amplitude and slope of the power-law, and $M_0 = 10^{13.5} \ \mathrm{M_{\odot}}$ is the pivot mass. \citet{Herle2026} already showed that a power-law is a good fit to the \pkg{FLAMINGO} data. However, at masses lower than $10^{13} \ \mathrm{M}_{\odot}$ and at masses higher than $\sim 10^{14.5}\ \mathrm{M}_{\odot}$, there is a deviation from the power-law at lower redshifts, shown in the bottom panel of Fig.~\ref{fig:a1_vs_mh_z}. This effect was already seen and described in \citet{Piras2018}. Using the Millenium and Millenium-XXL gravity-only simulations, they found that a power-law represents the alignment scaling with halo mass well, but deviates below $10^{13} \ \mathrm{M}_{\odot}$ and around $10^{14.5}-10^{15}\ \mathrm{M}_{\odot}$. At high masses, they detected a decreased alignment strength which they attributed to the haloes still being in the process of formation at $z=0$. Similarly, we see a decrease compared to the power-law by about 10-20 per cent for galaxies, for redshift 0 to 0.25. For the haloes of these objects we also see deviations from the power-law above this mass at the 5-10 per cent level above or below the power-law, which is also true in the gravity-only 2.8 Gpc box of \pkg{FLAMINGO}. At halo masses below $10^{13} \ \mathrm{M}_{\odot}$, \citet{Piras2018} saw an increase compared to the power-law, which they attributed to a different alignment mechanism being at play compared to the rest of the mass range. Since haloes in that mass range are infalling along filaments, they argued that they would have a stronger alignment amplitude than the power-law prediction. We also detect a deviation from the power-law, however, we find that it decreases by between 10-40 per cent. For the gravity-only 2.8 Gpc box, we see a residual of 10 per cent from the power-law at redshift 0, below a halo mass of $10^{13} \ \mathrm{M}_{\odot}$. Thus, this difference from the \citet{Piras2018} result cannot simply be a direct result of hydro-dynamics, and would require further exploration with other simulations with higher resolution. 

In light of these points, we restricted the mass range where we fit the power-law to between $10^{13}$ to $3 \times10^{14}\ \mathrm{M}_{\odot}$. The excluded regions are shown in grey in Fig.~\ref{fig:a1_vs_mh_z}. \citet{Kurita2021} also explored the mass and redshift dependence of the halo alignment amplitude in gravity-only simulations (see their figure 6), and we find the same qualitative trend as they did, that is that $A_1$ is larger for more massive haloes at the same redshift, and for the same halo mass, $A_1$ is larger for higher redshifts. 
 
\begin{figure}
    \centering
    \includegraphics[width=\columnwidth]{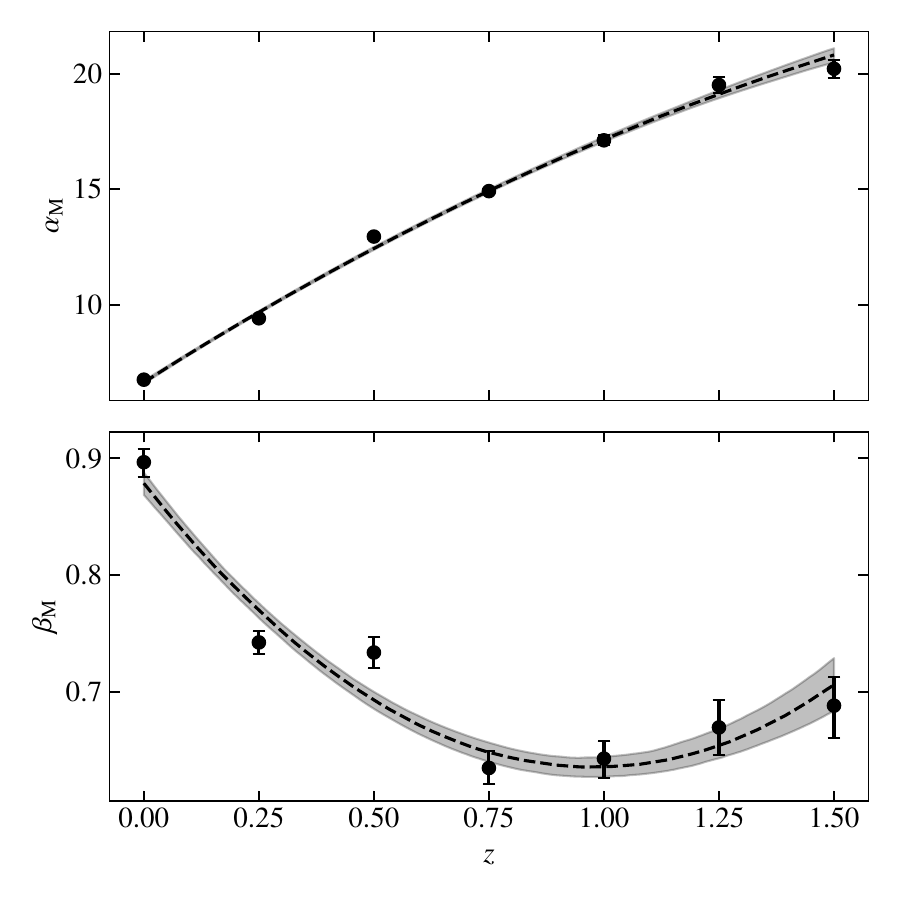}
    \caption{The variation of $\alpha_{\mathrm{M}}$ and $\beta_{\mathrm{M}}$ with redshift for LRG central galaxies. The amplitude of the alignment signal grows steadily with redshift while the slope drops with increasing redshift until $z\sim1$ and then flattens, with hints of increasing with redshift. The dashed line shows the best-fit quadratic function given by Eqns. \eqref{eqn:alpha} and \eqref{eqn:beta}.}
    \label{fig:alpha_beta_z}
\end{figure}

We show the variation of the best-fit $\alpha_{\mathrm{M}}$ and $\beta_{\mathrm{M}}$ with redshift in Fig.~\ref{fig:alpha_beta_z}. The amplitude of the power-law increases with increasing redshift, whereas the slope decreases up to $z \sim 1$ and then increases slightly. We fit quadratic functions to these data-points, resulting in 

\begin{equation}
\label{eqn:alpha}
    \alpha_{\mathrm{M}}(z) = -2.18 z^2 + 12.69z + 6.64,
\end{equation}

\begin{equation}
\label{eqn:beta}
    \beta_{\mathrm{M}}(z) = 0.26 z^2 -0.50z + 0.88.
\end{equation}

The fit coefficients are constrained to typical uncertainties of order 5 - 10 per cent when marginalised over all other parameters. Using Eqns. \eqref{eqn:mass}, \eqref{eqn:alpha} and \eqref{eqn:beta}, we have a mass-redshift model for the alignment amplitude. With these, we can predict the alignment amplitude for an LRG sample given the redshift and mean halo mass of the sample. We have, however, assumed perfect knowledge of the redshifts (i.e spectroscopic redshifts), and how the model parameters change for photometric redshifts is beyond the scope of this work. We can also recast the mass and redshift of an object with the peak height, $\nu$, which is a convenient way to reduce these two variables to just one, which is discussed in appendix~\ref{appendix:peak_height}.

\subsection{Tracking galaxies across snapshots}

\begin{figure*}
    \centering
    \includegraphics[scale=0.5]{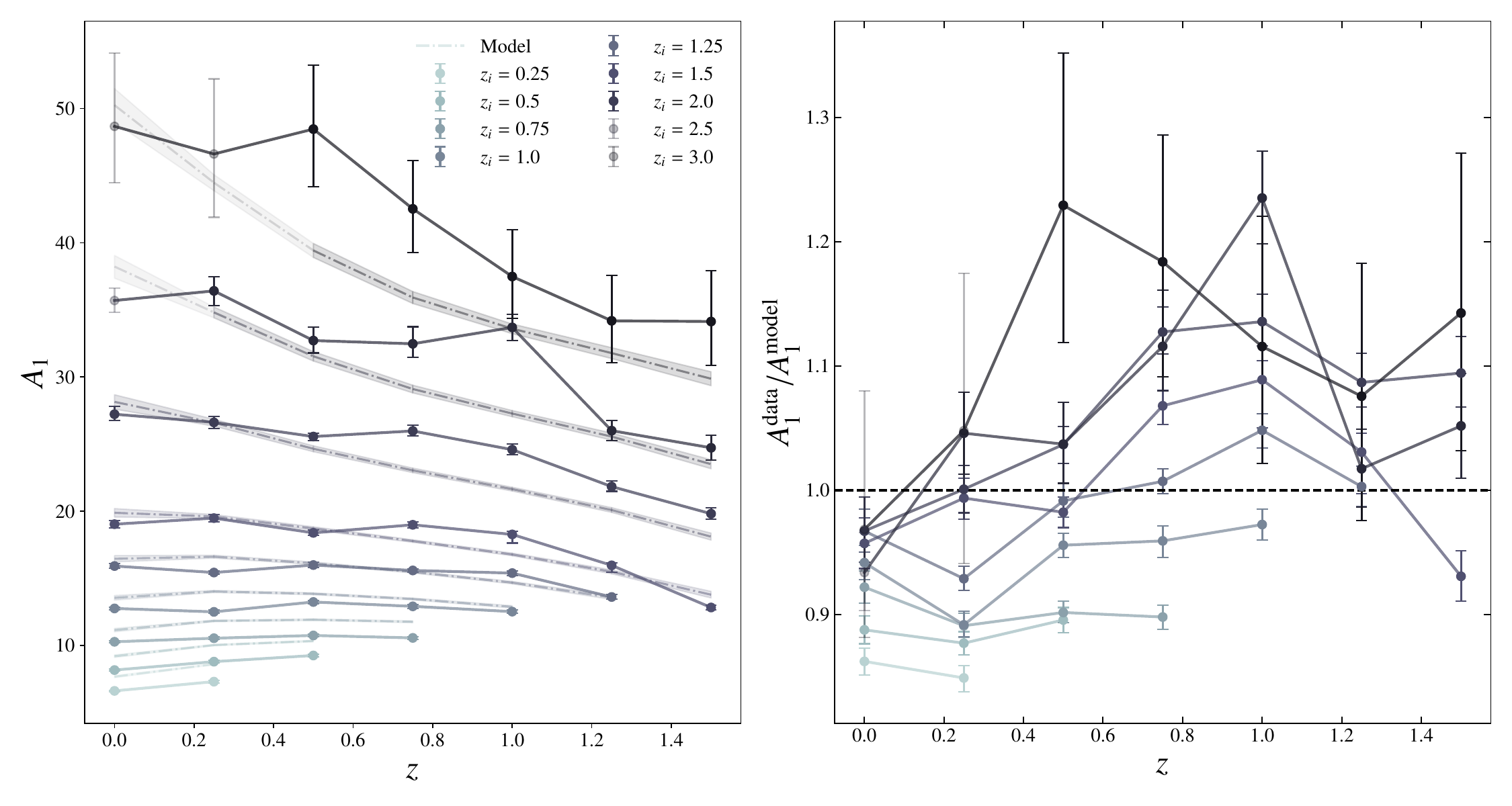}
    \caption{\textit{Left}: The variation of $A_1$ with redshift for central galaxies tracked from an initial redshift of $z_i$. Overplotted is the redshift-mass model from Eqns. \eqref{eqn:mass}, \eqref{eqn:alpha} and \eqref{eqn:beta} for each redshift given the halo mass of each sample. The points are lighter if the halo masses lie beyond the range of validity of this model. \textit{Right}: The ratio of the tracked $A_1$ and the model at each redshift from the left panel. The mass-redshift model introduced in this paper accounts for most of the variation in $A_1$ as there is a flattening and clustering around 1 after normalisation. There is, however, a systematic offset depending on the $z_i$, with galaxies tracked from higher $z_i$ having a higher $A^{\mathrm{data}}/A^{\mathrm{model}}$ than those tracked from lower $z_i$.}
    \label{fig:a1_tracked}
\end{figure*}

To isolate redshift evolution from simple mass growth, we track the same galaxy population across snapshots. From SOAP \citep{McGibbon2025} we have the \pkg{Track\_id} of each halo in the simulation. This property is unchanged for an object through snapshots (assuming it survives till that snapshot), and thus can be used to track objects across redshifts without having to use the full merger history of objects. Assigning particles to an object after a merger event in the HBT-HERONS scheme \citep{Moreno2025} requires deciding which subhalo will be considered the central. To do so, all subhaloes in the current FoF group are ranked according to their previous bound mass and only subhaloes whose bound masses are above 80 per cent of the most massive subhalo in the previous output are considered the new `central'. Given a situation where more than one object satisfies this condition, the object with the lowest specific orbital energy is chosen as the new central. For full details, we refer the reader to section 3.2.2 of \citet{Moreno2025}.

Using the \pkg{Track\_id}, we selected central galaxies that have more than 300 stellar particles, and whose dark matter haloes have a dark matter mass $> 10^{12} \ \mathrm{M}_{\odot}$ at some initial redshift $z_i$. The higher $z_i$ is, the smaller the resulting sample will be because very few galaxies have accreted the mass required to survive this cut. 

Once this sample is created at a specific $z_i$, it is tracked through different snapshots until redshift 0. At each snapshot, we remove the objects that are no longer centrals (when two centrals merge, by definition only one will become the new central of the merged object) and those that possess fewer than 300 stellar particles (about $\sim$ 0.5 per cent). At each redshift, the alignment amplitude is measured for the resulting sample. This exercise results in a variation of the alignment amplitude with redshift that is shown in the left panel of Fig.~\ref{fig:a1_tracked}. For $z_i > 1$, we see that $A_1$ increases with decreasing redshift, and for $z_i < 1$ there is a mild decrease. Using the mass-redshift model introduced in Eqns. \ref{eqn:mass}, \ref{eqn:alpha} and \ref{eqn:beta}, we can predict the alignment amplitude at each redshift given the halo mass of the objects in the sample. This prediction is shown by the dashed lines in the left panel of Fig.~\ref{fig:a1_tracked}.

To answer the question whether the alignment amplitude changes with redshift beyond the change in mass, we normalise the measured $A_1$ for the tracked snapshot at each redshift by the expected amplitude from the model. The ratio between the measured value, $A_{\mathrm{data}}$ and the model prediction, $A_{\mathrm{model}}$ is shown in the right panel of Fig.~\ref{fig:a1_tracked}. We see that the ratio is roughly flat with redshift, and around 1 for all $z_i$. This shows that our mass-redshift model accounts for most of the variation of the alignment amplitude with redshift and mass. Importantly, a systematic offset still remains, with $A^{\mathrm{data}}/A^{\mathrm{model}}$ being higher with increasing $z_i$. Essentially, central galaxies that satisfy our selection criteria earlier (i.e already formed at high redshift) will have a higher alignment amplitude than those galaxies that satisfy this criteria only later. 

We interpret this as another indication that assembly bias affects the alignment amplitude of an LRG central sample. Thus, accounting for the mass and redshift allows most of the variation of the alignment amplitude to be captured, but a change in amplitude of order 10 per cent at redshift 0 is still unaccounted for. As we showed in Fig.~\ref{fig:a1_vs_mh_zform}, the effect of assembly bias is a shift in the amplitude of the alignment signal. Therefore, each sample with different $z_i$ will have a different offset due to assembly bias. In principle, one could also account for this shift in the mass-redshift model, by making $\alpha_{\mathrm{M}}$ also a function of $z_{\mathrm{f}}$. This would however, require further binning the sample at each redshift into bins of $z_{\mathrm{f}}$ in addition to halo mass, for which the \pkg{FLAMINGO} sample size is too low to obtain meaningful results beyond very low redshifts. 

\subsection{Comparison with previous works}

\citet{Chisari2016} studied the redshift evolution of galaxies in the Horizon-AGN simulations, and found that the alignment signal changes from radial to tangential with increasing redshift. Although disks dominate the high redshift signal (thereby decreasing the radial alignment), the evolution of galaxies from disks to ellipticals alone was insufficient to explain their result, requiring also an inherent change in the alignment signal. For a sample of elliptical galaxies, an increase of the alignment strength with redshift was seen, consistent with our findings. The increase with mass and redshift was also noted by \citet{Zjupa2022} in IllustrisTNG. 

The redshift evolution of galaxy alignments has been studied by \citet{Bhowmick2020} using the \pkg{MassiveBlackII} hydro-dynamic simulation. By analysing merger trees of this simulation, they compared how the shapes of dark matter and stellar components change with redshift, as well as the misalignment angle between galaxies and their haloes. In addition, \citet{Bate2020} tracked the progenitors of ellipticals at redshift 0.06 of the Horizon-AGN simulations, and found that they go from having no alignment with the tidal field at redshift 3 to having a significant alignment at redshift 1, and then plateauing out after $z\sim0.5$. This result is similar to the $z_i = 3.0$ curve in the left panel of Fig.~\ref{fig:a1_tracked}, where we also see an increase in the amplitude from higher redshifts to lower redshifts, and then a constant amplitude from redshift 0.5. \citet{Bate2020} also noted that the increase in the amplitude with decreasing redshift was accompanied by an increase in the mean mass of the sample. 

In this work, we improve upon these results by using a sample of LRGs with minimal redshift evolution in the galaxy disk fractions (see appendix~\ref{appendix:disk_fraction}), so that the inherent change in the IA amplitude can be studied. We further studied the change in the amplitude beyond the change associated with the mass accretion of the sample, which we accounted for with our mass-redshift model in the right panel of Fig.~\ref{fig:a1_tracked}. 

\section{Discussion and conclusions}
\label{sect:disc_conc}

The intrinsic alignment of galaxies is one of the main astrophysical systematics in weak lensing shear analyses. Since IA cannot be predicted from first principles, hydro-dynamic simulations are instrumental in providing guidance for the modeling of the alignment signal in surveys like \textit{Euclid} and LSST. One key aspect of IA that is not yet understood in the literature is its redshift evolution. Moreover, the amplitude is often only considered to be a function of halo mass.

In this paper, we used the 2.8 Gpc box of the \pkg{FLAMINGO} simulation to explore the effect of formation redshift and the redshift evolution of the alignment signal. The dependence of halo properties beyond mass, dubbed assembly bias, has been studied extensively in the context of galaxy clustering \citep{Sheth2004, Gao2005}. The impact of assembly bias on alignments, however, has not yet been studied in detail in the literature. In this work, we showed that the earlier a halo or galaxy formed, the more strongly aligned it is. Essentially, galaxies with the same halo mass can have different values of $A_1$, with galaxies that formed earlier having a systematically higher amplitude (Fig.~\ref{fig:a1_vs_mh_zform}).

This result has important implications for the modeling of alignments in observations. For haloes of the same mass, the higher the formation redshift of a sample of galaxies, the stronger their alignment is.  The slope of the power-law with mass remains unchanged with changes in $z_\mathrm{f}$, and the effect of assembly bias only changes the amplitude of the signal. Since the alignment amplitude is a free parameter in cosmological analyses anyway, our results imply that there is no need to account directly for assembly bias in our IA models. It does, however, imply that the prior range on the amplitude cannot be arbitrarily small, as we would need to account for the spread in the alignment amplitude at a given halo mass. 

In the case of HOD models, the impact of assembly bias has important consequences. Typically, HOD-based models assign halo properties based on halo mass which would not capture this additional dependence on halo formation history \citep[although using color information as well could be useful, ][for example]{Castander2025}. This implies the need for more sophisticated HOD models that account for the impact of assembly bias on IA, and also highlights the importance of fully hydrodynamic simulations for IA studies. 

Observationally, detecting the effect of assembly bias on IA is more difficult, as the formation redshift is not directly observable. Observational proxies of the formation redshift, however, like the Sérsic index, $n_\mathrm{S}$, could be used to explore this in data. Interestingly, \citet{Georgiou2025} used the KiDS-1000 bright sample to measure the alignment amplitude for red galaxies split by $n_\mathrm{S}$. They found that red galaxies with $n_\mathrm{S} > 4$ had an $A_1$ of $5.24\pm0.85$ and red galaxies with $n_\mathrm{S} < 4$ had a lower amplitude of $1.12\pm0.88$. This means that galaxies that have a lower $n_\mathrm{S}$ have a lower alignment strength, and since typically a lower $n_\mathrm{S}$ corresponds to a lower concentration, more concentrated objects tend to align stronger. Since concentration and formation redshift are strongly correlated (Fig.~\ref{fig:c_vs_zform}), this detection could potentially be interpreted as the effect of assembly bias that we explored in this work \citep[see also][]{Laigle2025}. However, contamination from red spirals, which are essentially disk galaxies affected by dust attenuation, or that posses a dominant bulge, would cause them to end up in a red sample and then pull the alignment amplitude down, complicating the interpretation. Moreover, the difference between the $n_\mathrm{S} > 4$ and $n_\mathrm{S} < 4$ is only 2$\sigma$. Repeating this kind of analysis using the larger sample with better resolved morphologies to remove the red spirals with \textit{Euclid} could be a future observational extension of this work.

Apart from the Sérsic index, the impact of assembly bias on IA could also be studied using other proxies of formation time. For example, the sSFR or the stellar age could be used to separate a galaxy sample into older and younger populations \citep{Wang2013, Lacerna2014}, which could be used to place a minimum prior width for the alignment amplitude in cosmological analyses. In simulations, one could look for departures from the co-evolution relations by fitting higher order intrinsic alignment models in analogy to the work done by \citet{Lazeyras2021} in the case of galaxy clustering.

We further explored how the alignment signal evolves with redshift. Simply making the same mass cut at different redshifts would not account for the evolution of the mass with redshift, as haloes typically accrete mass over time. We accounted for this mass dependence by looking at the evolution of the power-law parameters, $\alpha_{\mathrm{M}}$ and $\beta_{\mathrm{M}}$ with redshift. We did this by first fitting power-laws to the relation between the alignment amplitude and halo mass at different redshifts (Fig.~\ref{fig:a1_vs_mh_z}). We found deviations from the power-law outside the mass range of $10^{13} - 3\times10^{14} \ \mathrm{M}_{\odot}$, in qualitative agreement with \citet{Piras2018} at high masses. At low masses, we see a decrease in both the hydro- and gravity only simulations, whereas they observed a positive residual compared to the power-law. This requires further exploration with higher resolution hydro-simulations, for example with the \pkg{COLIBRE} simulations \citep{Schaye2025, Chaikin2025a}.

We provide empirical fits to the variation of the power-law parameters with redshift (Fig.~\ref{fig:alpha_beta_z}), and Eqns. \eqref{eqn:mass}, \eqref{eqn:alpha} and \eqref{eqn:beta} together constitute a mass-redshift IA model based on the \pkg{FLAMINGO} simulations. The power-law amplitude increases with redshift, whereas the slope decreases until $z\sim1$ and then flattens, with mild indications of increasing with redshift. This potentially indicates that the alignment mechanism at play changes around this stage of the universe's evolution, as most structures have already formed by this redshift. 

One caveat though is that given the resolution of the \pkg{FLAMINGO} simulations, we do not properly capture the evolution of galaxies with redshift. The disk fraction (defined as the fraction of our sample with $|V / \sigma | > 1$) remains below 1 per cent out to $z\sim1.5$ (see appendix~\ref{appendix:disk_fraction} for details). The minimal evolution of the disk fraction is due to the fact that the resolution of our simulation is insufficient to capture the flattening of ellipticals into disks. Thus, our results do not account for disks or the evolution of ellipticals into disks, and our mass-redshift model can strictly only be used for a pure LRG sample. However, in the mass range considered here, this evolution will be minimal. We also note that our model is only valid for halo masses in the range of $10^{13}-10^{14.5} \ \mathrm{M}_{\odot}$ and up to $z\sim1.5$.

With this resolution limitation in mind, we tracked galaxies across snapshots. We found that alignment amplitude increases with redshift for galaxies tracked from higher initial redshifts, whereas those tracked from lower $z_i$ remain fairly constant. With time, the mass of these galaxies increases due to accretion, and to account for this mass growth, we normalised the amplitudes by the prediction from our mass-redshift model. The normalised curves (Fig.~\ref{fig:a1_tracked}) are roughly flat, showing that our model captures most of the mass and redshift evolution of the IA signal. Beyond this flattening, we observed a higher ratio of $A^{\mathrm{data}}/A^{\mathrm{model}}$ for the galaxies tracked from higher $z_i$, which again shows the impact of assembly bias on the alignment amplitude. 

In conclusion, we showed that assembly bias impacts IA. We also introduced a mass-redshift evolution model for the alignment amplitude. The main limitation in this work is the resolution of the simulation not being sufficient to resolve disks and galaxy evolution of our sample. Future work using the \pkg{COLIBRE} simulations \citep{Schaye2025, Chaikin2025a} will be instrumental in pinning down the redshift evolution of the alignment signal. 

\begin{acknowledgements}
AH thanks Casper Vedder, Benjamin Joachimi and Boryana Hadzhiyska for valuable discussions. AH acknowledges support by NWO through the Dark Universe Science Collaboration (OCENW.XL21.XL21.025). 

NEC acknowledges support from the project ``A rising tide: Galaxy intrinsic alignments as a new probe of cosmology and galaxy evolution'' (with project number VI.Vidi.203.011) of the Talent programme Vidi which is (partly) financed by the Dutch Research Council (NWO). 

HH and DN acknowledge funding from the European Research Council (ERC) under the European Union's Horizon 2020 research and innovation program (Grant agreement No. 101053992).

We acknowledge the Virgo Consortium for making their simulation data available. The FLAMINGO simulations were performed using the Durham Memory Intensive system managed by the Institute for Computational Cosmology on behalf of the STFC DiRAC facility (www.dirac.ac.uk).

This work employed the packages \pkg{NUMPY} \citep{Harris2020}, \pkg{MATPLOTLIB} \citep{Hunter2007}, \pkg{SCIPY} \citep{Scipy2020}, \pkg{SWIFTsimIO} \citep{Borrow2020} and CCL \citep{Chisari2019}.

This work benefited from the resources and support provided by the echo-IA collaboration (https://echo-ia.org).

\end{acknowledgements}
\bibliographystyle{aa} 
\bibliography{main}

\begin{appendix}
\section{Disk fraction}
\label{appendix:disk_fraction}

\begin{figure}
    \centering
    \includegraphics[width=\columnwidth]{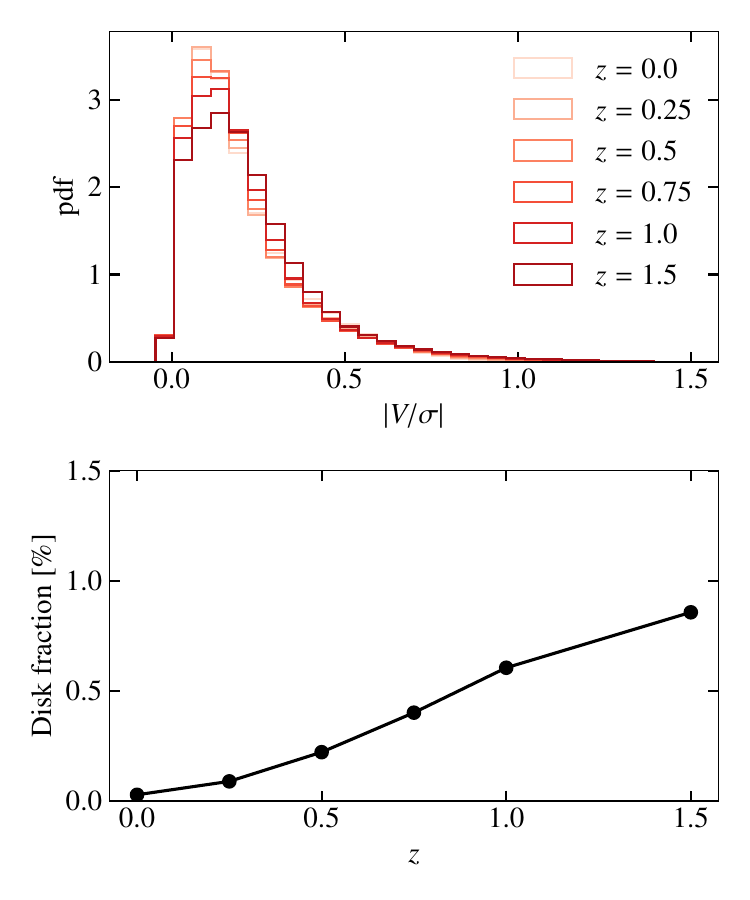}
    \caption{\textit{Top}: The distribution of $|V/\sigma|$ for our sample at different redshifts. \textit{Bottom}: the disk fraction of our sample assuming disk galaxies have $| V/\sigma| > 1$.}
    \label{fig:v_sig}
\end{figure}

Galaxies can be split into early, red galaxies (LRGs) or blue, starforming disky galaxies. This can be done based on their morphology, colour or kinematics. For a comparison of how different splits affect the measured IA signal, we refer the reader to \citet{vanHeukelum2025b}. In this appendix, we used the kinematics of the galaxies to measure the fraction of galaxies in our sample that are disky. We measure the absolute value of the ratio between the rotational velocity with the velocity dispersion of the stellar particles, $| V/\sigma|$. For a detailed derivation of this parameter, we refer the reader to Eqns. 18-21 of \citet{vanHeukelum2025b}.

Galaxies that are pressure supported (LRGs) will have a $| V/\sigma| < 1$, and we use this threshold to define disk galaxies. The distribution of $| V/\sigma|$ at various redshifts in the simulation are shown in the top panel of Fig.~\ref{fig:v_sig}. We see minimal change in the distributions of $| V/\sigma|$, which is reflected also in the disk fraction shown in the bottom panel. This is likely a direct result of insufficient mass resolution of the simulation used in this work. The disk fraction goes from almost 0 per cent at $z=0$, to about 1 per cent at $z=1.5$. This is a minimal change in disk fraction, which implies that our sample is always an LRG sample with minimal evolution from rotational to pressure supported galaxies.

\section{$A_1$ as a function of peak height}
\label{appendix:peak_height}
\begin{figure}
    \centering
    \includegraphics[width=\columnwidth]{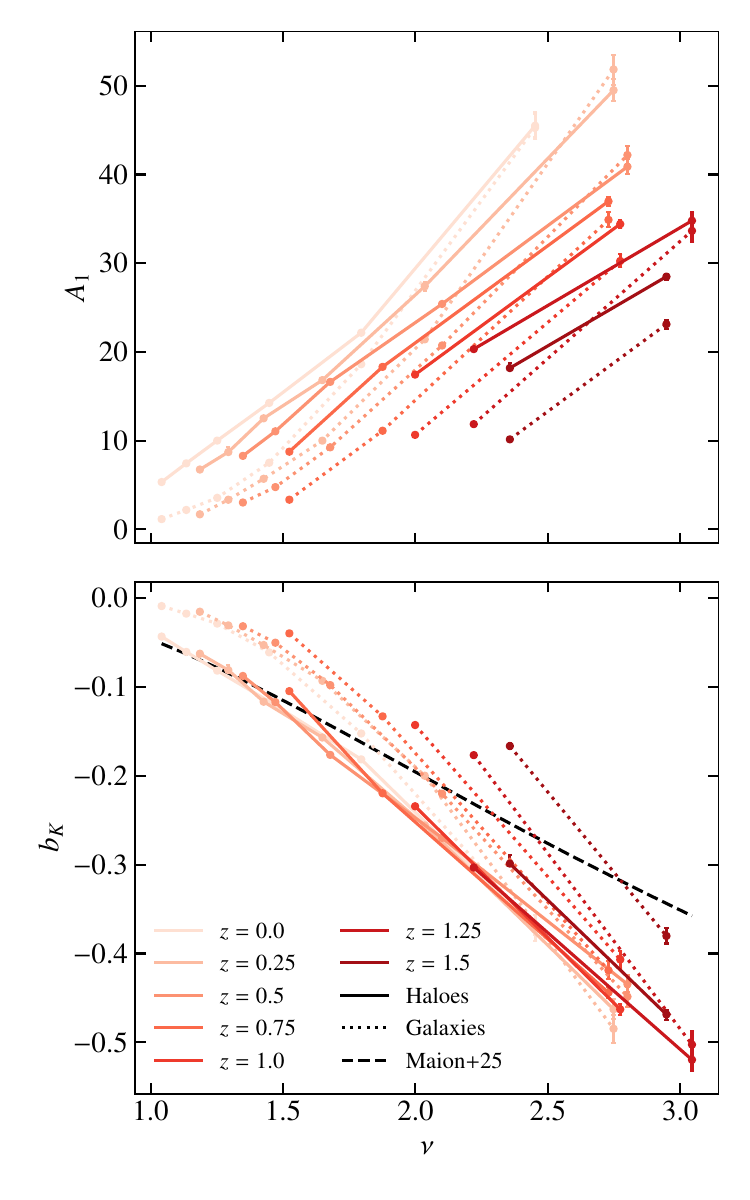}
    \caption{The alignment amplitude, $A_1$ and the shape bias parameter, $b_k$ as functions of the peak height for different redshifts, for both haloes and galaxies. Over plotted is a fit from \citet{Maion2025}.}
    \label{fig:A_vs_nu}
\end{figure}

Fig.~\ref{fig:a1_vs_mh_z} showed the alignment amplitude as a function of halo mass for different redshifts. We can attempt to recast this relation between $A_1$ and the mass and redshift as a relation between $A_1$ and the peak height, $\nu$, which is defined as:

\begin{equation}
    \nu = \delta_{\mathrm{c}}/\sigma(M, z)
\end{equation}
where $\delta_{\mathrm{c}}$ is the linearly extrapolated critical collapse threshold of 1.686, and $\sigma$ is the r.m.s density fluctuation smoothed on mass scale $M$ at a given redshift $z$. Recasting in this way is a more natural representation of the relation between $A_1$ and the mass and redshift from a perturbation theory perspective. 

This enables us to compare against the results of \citet{Maion2025}, in particular with their figure 4 (left-most panel) and figure 6. In the top panel of Fig.~\ref{fig:A_vs_nu} we show the alignment amplitude as a function of peak height for each redshift, for both galaxies and haloes. We implemented this using \pkg{pyccl} \citep{Chisari2019}. We computed the shape-bias parameter from the alignment amplitude using $b_K = - 2 A_1 C_1 \rho_{\mathrm{crit}}\Omega_{\mathrm{m}, 0} D(z)^{-1}$, with $C_1\rho_{\mathrm{crit}}$ fixed to 0.0134, and show it as a function of peak-height in the bottom-panel. As noted in the left-panel of figure 4 of \citet{Maion2025}, the points all lie on the same line for haloes. Our result is in qualitative agreement with theirs despite the lack of hydro-dynamics in their simulations. In contrast, this trend does not hold for galaxies. In dashed, we show the fit from that paper for reference, although we do not expect a match given that our simulation was run for a different cosmology and with hydro-dynamics.

\begin{figure}
    \centering
    \includegraphics[width=\columnwidth]{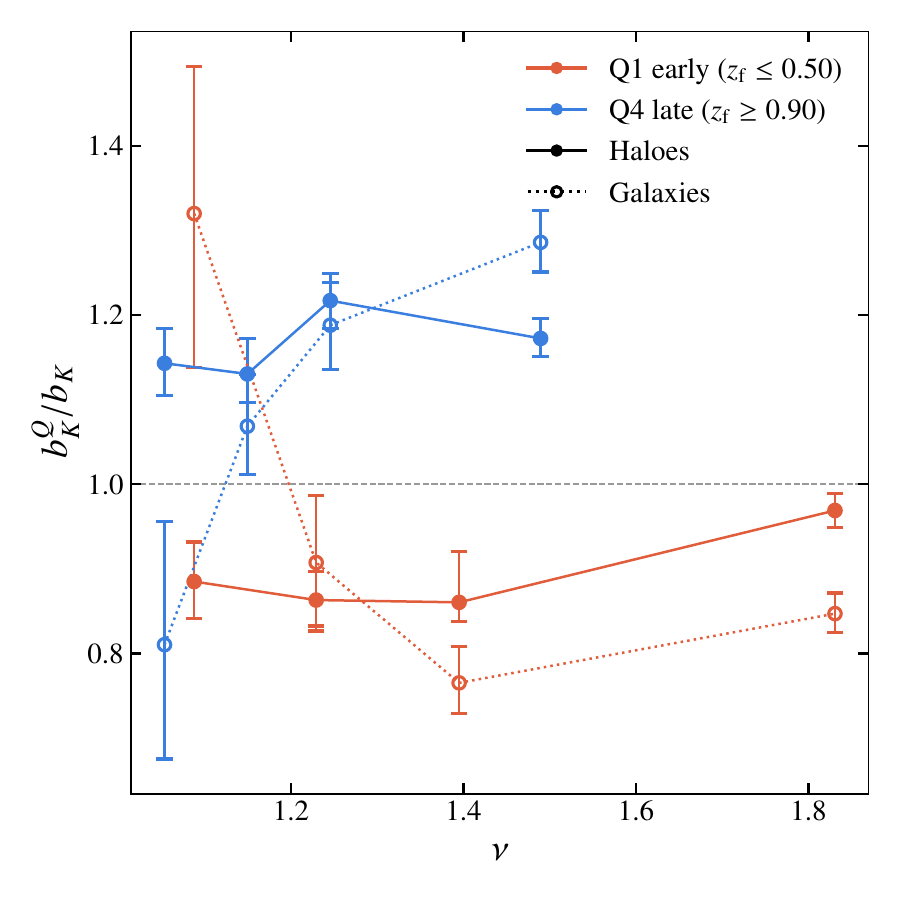}
    \caption{$b_k$ as a function of peak-height $\nu$ for the earliest and latest forming objects in the simulation, normalised by the $b_k$ of the full sample, for both galaxies and haloes. A clear secondary dependence beyond peak-height is seen, with earlier forming haloes having a lower $b_K^Q/b_k$.}
    \label{fig:bk_nu_quartiles}
\end{figure}

Figure 6 of \citet{Maion2025} showed a clear secondary dependence of the shape-bias parameter on halo spin, by splitting the sample into quartiles in the distribution of spins. They found that at low $\nu$, lower halo spin haloes have a higher $b_k$ than the full sample, which is inverted at some characteristic peak-height of $\nu \sim1.75$. Similarly, we split our sample into the lowest and highest quartiles in formation redshift (Q1 and Q4), and calculated $b_K$ as a function of $\nu$. This was then normalised by the full sample, and the results are shown in Fig.~\ref{fig:bk_nu_quartiles}. For haloes, we see a clear separation of the two populations, with later forming objects (Q4) having a systematically higher $b_k$ than the earlier forming objects. This can be understood easily from Fig.~\ref{fig:a1_vs_mh_zform} as earlier forming haloes have a higher alignment amplitude. For the galaxies, there is an inversion at lower $\nu$ after which they too follow the same trend as the haloes. Thus, we find a clear detection of assembly bias on the alignment amplitude.

\end{appendix}

\end{document}